%% file: main.tex
\documentclass{article}
\usepackage[T1]{fontenc}
\usepackage{microtype}
\usepackage[utf8]{inputenc}
\usepackage{ismir}
\usepackage{amsmath,cite,url}
\usepackage{graphicx}
\usepackage{color}
\usepackage{booktabs}

\usepackage{amssymb}
\usepackage{algorithm}
\usepackage[noend]{algpseudocode}

\usepackage{tikz}

\usetikzlibrary{shapes.geometric, arrows.meta, positioning, calc}

\usepackage{xcolor}

\usepackage[nolist]{acronym}
\begin{acronym}
\acro{stft}[STFT]{Short-Time Fourier Transform}
\acro{istft}[iSTFT]{Inverse Short-Time Fourier Transform}
\acro{dnn}[DNN]{deep neural network}
\acro{pesq}[PESQ]{Perceptual Evaluation of Speech Quality}
\acro{polqa}[POLQA]{perceptual objective listening quality analysis}
\acro{wpe}[WPE]{weighted prediction error}
\acro{psd}[PSD]{power spectral density}
\acro{rir}[RIR]{room impulse response}
\acro{hrir}[HRIR]{head-related impulse response}
\acro{brir}[BRIR]{binaural room impulse response}
\acro{hrtf}[HRTF]{head-related transfer function}
\acro{snr}[SNR]{signal-to-noise ratio}
\acro{lstm}[LSTM]{long short-term memory}
\acro{polqa}[POLQA]{Perceptual Objective Listening Quality Analysis}
\acro{sdr}[SDR]{signal-to-distortion ratio}
\acro{cnns}[CNNs]{Convolutional Neural Networks}
\acro{vae}[VAE]{variational auto-encoder}
\acro{gan}[GAN]{generative adversarial network}
\acro{tf}[T-F]{time-frequency}
\acro{sde}[SDE]{stochastic differential equation}
\acro{ode}[ODE]{ordinary differential equation}
\acro{drr}[DRR]{direct to reverberant ratio}
\acro{lsd}[LSD]{log spectral distance}
\acro{sisdr}[SI-SDR]{scale-invariant signal to distortion ratio}
\acro{mos}[MOS]{mean opinion score}
\acro{map}[MAP]{maximum a posteriori}
\acro{sde}[SDE]{stochastic differential equation}
\acro{ode}[ODE]{ordinary differential equation}
\acro{dps}[DPS]{diffusion posterior sampling}
\acro{tfrs}[TFRs]{Time-Frequency Representations}
\acro{cqt}[C$Q$T]{Constant-$Q$ Transform}
\acro{ldm}[LDM]{Latent Diffusion Model}
\end{acronym}

\title{Music Inpainting with Similarity-Guided Latent Diffusion}

\multauthor
  {Sean Turland$^{1,\hspace{1pt}2}$ \hspace{1cm} Eloi Moliner$^1$ \hspace{1cm}  Vesa Välimäki$^1$ }
  {
  $^1$ Acoustic Lab, Dept. Information and Communications Engineering, Aalto University, Finland\\
  $^2$ Acoustics and Audio Group, University of Edinburgh, United Kingdom \\
  {\tt\small sean.turland@ed.ac.uk, eloi.moliner@aalto.fi,
  vesa.valimaki@aalto.fi}
  }

\def\authorname{S. Turland, E. Moliner, and V. Välimäki}

\begin{document}

\maketitle

\begin{abstract}

\noindent We propose a music inpainting technique combining similarity matching and a diffusion-based generative model to replace large segments of audio recordings with suitable alternatives.
The proposed hybrid method, Similarity-Guided Diffusion Posterior Sampling (SimDPS), retrieves candidate replacement segments from the wider signal based on contextual similarity
and uses them to guide the diffusion process toward coherent reconstructions.
Being model-agnostic, the framework is compatible with any pre-trained diffusion model operating in either the waveform or latent domain.
Objective evaluations demonstrate that SimDPS outperforms widely adopted RePaint and DPS methods across acoustic features and neural-embedding distance metrics.
Subjective evaluations on multi-genre music datasets confirm that SimDPS enhances perceptual plausibility compared to previous diffusion-based methods and demonstrates robustness to poorly matched similarity candidates. These findings pave the way for improved music restoration and context-aware audio editing.

\end{abstract}

\section{Introduction}\label{sec:introduction}

Audio inpainting \cite{adler_audio_2012} aims to reconstruct missing segments of a signal with perceptually plausible content.
It has applications in packet-loss concealment for data transmission \cite{sanneck_new_1996}, restoration of recordings stored on degraded physical media \cite{moliner_diffusion-based_2024}, and correction of errors in post-production audio editing \cite{levy_controllable_2023}.
Distinct applications place different demands on the duration of the missing material: ranging from gaps of a few milliseconds in packet-loss concealment to segments lasting several seconds in audio restoration and editing.

This paper focuses on the long-gap regime, which is particularly challenging because it requires the model to go beyond local signal continuity and preserve long-range musical coherence.
At this scale, the missing region may contain an entire phrase or structural unit, such as a chorus or motif, that is central to the track's identity.
While classical methods (e.g., autoregressive modelling
\cite{janssen_adaptive_1986,kauppinen_audio_2002,esquef_interpolation_2003}
or time-frequency sparsity
\cite{mokry_introducing_2019, taubock2020dictionary}) effectively handle short gaps, their reliance on signal stationarity makes them unsuitable for multi-second gaps.

Two main paradigms currently address long-gap inpainting.
The first is self-similarity search, which exploits the redundancy and structural recurrence of musical signals \cite{foote1999visualizing} by retrieving candidate segments from observed portions of the same track \cite{martin2011exemplar, perraudin_inpainting_2018}.
This paradigm is particularly effective for music with strong repetition, where missing segments can be replaced by near-verbatim copies without breaking stylistic consistency.
Its main limitation is that it remains deterministic: once a sufficiently close match is unavailable, the method cannot adapt the retrieved material to the local details of the target context.

A second paradigm is based on generative modelling, where masked regions are reconstructed by sampling from a learned probabilistic prior, such as a diffusion model \cite{sohl-dickstein_deep_2015, ho_denoising_2020}.
Diffusion models have been previously used for audio inpainting, either by training with paired original and masked signals \cite{kong2025a2sb}, or employing training-free techniques, such as posterior sampling approaches \cite{moliner2023solving, moliner_diffusion-based_2024},
or initial noise optimisation \cite{novack_ditto_2024, ben2024d}.
\acp{ldm} have also been used for audio inpainting \cite{liu2023audioldm, wang2023audit}, and are attractive because compressed representations can model multi-second context efficiently.
Unlike deterministic retrieval, these generative approaches can synthesise novel content that blends with the surrounding audio.

However, generative inpainting methods are prone to identity drift as gap duration increases.
Even when the generated audio is locally smooth and acoustically plausible, it may fail to preserve track-specific attributes such as melody or rhythmic placement.
A way to mitigate this drift is conditioning the generation process on auxiliary controls, such as piano rolls \cite{liu2023maid} or text descriptions \cite{wang2023audit, liu2023audioldm, novack_ditto_2024}.
Yet such side information is rarely available in practice.

We propose \textit{SimDPS}, a novel hybrid method that uses similarity-based retrieval to provide posterior guidance for diffusion models.
The method combines the structural reliability of self-similarity search with the expressivity of generative modelling, allowing the long-gap reconstruction to remain contextually grounded while still adapting to the local musical details.
The search procedure produces multiple weighted candidates, around which we design an aggregated spectral-latent inference scheme.
Crucially, our approach is domain-agnostic, serving as a plug-in for both latent and waveform-based architectures.\footnote{Source code and audio examples are provided at the following
URL: \texttt{https://s-turland.github.io/SimDPSInpainting/}}

\vspace{-3pt}
\section{Background}
\vspace{-2pt}

Audio inpainting can be formulated as a linear inverse problem.
Let $\mathbf{X} \in \mathbb{R}^{N \times C}$ denote the clean representation to be reconstructed, where $N$ is the number of temporal positions and $C$ is the feature dimension.
In the single-channel waveform domain, $C=1$, whereas in a latent representation produced by an audio autoencoder we may have $C$ channels.
The degraded observation satisfies
\begin{equation}
    \mathbf{Y} = \mathbf{M}(\mathbf{X} + \boldsymbol{\varepsilon}_Y),
    \label{eq:inverse}
\end{equation}
where $\mathbf{Y} \in \mathbb{R}^{N \times C}$ denotes the observation, the temporal mask $\mathbf{M} \in \mathbb{R}^{N\times N}$ is a binary operator that preserves the observed time indices and zeros out the missing interval, and the same mask is applied to every feature channel.
The observation uncertainty is modelled as Gaussian noise, $\boldsymbol{\varepsilon}_Y \sim \mathcal{N}(\mathbf{0}, \sigma_Y^2 \mathbf{I})$, where $\sigma_Y$ controls the noise standard deviation.
The goal of audio inpainting is therefore to recover the original representation $\mathbf{X}$.

The temporal mask $\mathbf{M}$ is diagonal and binary, with elements $\mathbf{M}_{ij} = m_i \delta_{ij}$, where $\delta_{ij}$ is the Kronecker delta and $m_i \in \{0, 1\}$ indicates whether temporal index $i$ is observed.
As we focus on compact gaps, the mask is restricted to a missing interval $[t_s, t_e]$ of length $L = t_e - t_s + 1$, with $m_i = 0$ for $t_s \leq i \leq t_e$ and $m_i = 1$ otherwise.

Because $\mathbf{M}$ is non-injective, the forward model has a non-trivial null space.
The null space of $\mathbf{M}$ is
$\{ \mathbf{Z} \in \mathbb{R}^{N \times C} \,:\, \mathbf{M}\mathbf{Z} = \mathbf{0} \}$,
corresponding to content supported entirely on the missing interval.
Audio inpainting is therefore ill-posed and requires prior information or structural assumptions to recover a perceptually consistent solution.

\vspace{-3pt}
\subsection{Diffusion Probabilistic Models}
\vspace{-2pt}

Diffusion probabilistic models are generative models that can sample from complex distributions and have attracted substantial attention in recent years \cite{song_score-based_2021, karras2022elucidating}.
They can be used to model the distribution $p_X$ of clean signals, either directly in waveform space or indirectly in the latent space of an autoencoder, from a finite dataset of examples.
This learned distribution can then serve as a data-driven prior over plausible clean reconstructions for ill-posed inverse problems such as audio inpainting \cite{moliner_diffusion-based_2024}.

More specifically, diffusion models define an iterative mapping, controlled by a diffusion-time variable $\tau$, that transforms samples from a tractable (usually Gaussian) prior distribution $p_T \sim \mathcal{N}(\mathbf{0},\sigma^2(T)\mathbf{I})$ at $\tau = T$ into samples from the data distribution $p_X$ at $\tau = 0$.
Following \cite{karras2022elucidating}, this mapping can be described by the \textit{probability-flow ordinary differential equation}, which defines a deterministic, bijective transformation:
\begin{equation}
    d\mathbf{X} = - \dot \sigma(\tau)\sigma(\tau) \, \nabla_{\mathbf{X}} \log p_{\tau}(\mathbf{X}; \sigma(\tau)) \, d\tau \, .
\label{eq:probflow}
\end{equation}
Here, $\dot\sigma(\tau)$ denotes the derivative of $\sigma(\tau)$ with respect to $\tau$.
Following \cite{karras2022elucidating}, we set $\sigma(\tau) = \tau$.
Diffusion models approximate the intractable score function with a neural network
$s_\theta(\mathbf{X}_\tau, \tau) \approx \nabla_{\mathbf{X}} \log p_{\tau}(\mathbf{X}; \sigma(\tau))$, which is trained with a denoising score matching objective \cite{vincent2011connection}.
In practice, new samples from $p_X$ are obtained by initialising $\mathbf{X}_T \sim p_T$ and numerically solving the probability-flow equation in \eqref{eq:probflow}.

An important property of the score function under Gaussian corruption is that it can be expressed in terms of the minimum mean-squared-error (MMSE) estimator of the clean sample $\mathbf{X}_0$ given $\mathbf{X}_\tau$.
By Tweedie's formula \cite{efron2011tweedie},
\begin{equation}
\hspace{-4pt}
    \nabla_{\mathbf{X}} \log p_\tau(\mathbf{X}; \sigma(\tau))
\hspace{-2pt}
    =
\hspace{-2pt}
    \tfrac{1}
        {\sigma^2(\tau)}
        (\mathbb{E}[\mathbf{X}_0
\hspace{-2pt}
        \mid
\hspace{-2pt}
        \mathbf{X}_\tau; \sigma(\tau)]
\hspace{-1pt}
        -
\hspace{-1pt}
        \mathbf{X}_\tau) .
    \label{eq:tweedie}
\end{equation}
This connection shows that the score network can be interpreted as a Gaussian denoiser:
$\hat{\mathbf{X}}_0(\mathbf{X}_\tau)= \mathbf{X}_\tau+ \sigma^2(\tau) s_\theta(\mathbf{X}_\tau, \tau) \approx \mathbb{E}[\mathbf{X}_0 \mid \mathbf{X}_\tau; \sigma(\tau)]$.

\vspace{-3pt}
\subsection{Diffusion Posterior Sampling}
\vspace{-2pt}

An appealing feature of diffusion models is that they can be adapted to solve inverse problems without task-specific retraining.
This work builds on Diffusion Posterior Sampling (DPS) \cite{chung2023diffusion} to solve the audio inpainting inverse problem \cite{moliner_diffusion-based_2024}.
Specifically, DPS steers sampling toward the conditional posterior $p(\mathbf{X}\mid\mathbf{Y})$ by replacing the prior score in \eqref{eq:probflow} with an approximation to the posterior score.
By Bayes' rule, the posterior score decomposes as
\begin{equation}
\hspace{-2pt}
\underbrace{\nabla_{\mathbf{X}_\tau}
\hspace{-2pt}
 \log p_\tau
\hspace{-1pt}
 (\mathbf{X}_\tau
\hspace{-2.5pt}
\mid
\hspace{-2.5pt}
\mathbf{Y}
\hspace{-1pt}
)}_{\text{Posterior Score}}
\hspace{-3pt}
=
\hspace{-3pt}
\underbrace{\nabla_{\mathbf{X}_\tau}
\hspace{-2pt}
\log p_\tau
\hspace{-1pt}
(
\hspace{-1pt}
    \mathbf{X}_\tau
\hspace{-1pt}
)}_{\text{Prior Score}}
\hspace{-1.5pt}
 +
\hspace{-1.5pt}
 \underbrace{\nabla_{\mathbf{X}_\tau}
\hspace{-2pt}
 \log p_\tau
\hspace{-1pt}
 (\mathbf{Y}
\hspace{-2.5pt}
 \mid
\hspace{-2.5pt}
  \mathbf{X}_\tau
\hspace{-1pt}
  )}_{\text{Likelihood Score}} \, .
\label{eq:bayes_score_decomp}
\end{equation}
While the prior score is approximated by the trained diffusion model $s_\theta(\mathbf{X}_\tau, \tau)$, the likelihood score is typically approximated from the assumed observation model.

Following \cite{chung2023diffusion}, we introduce a Gaussian surrogate likelihood on the observed signal, centred at the denoised estimate produced by the score model and using the observation-uncertainty scale parameter $\sigma_Y^2$.
This yields, up to a constant factor absorbed into the guidance scaling, the likelihood-score approximation
\begin{equation}
\hspace{-6pt}
\nabla_{\mathbf{X}_\tau}
\hspace{-1.5pt}
 \log p_\tau(\mathbf{Y}
\hspace{-1.5pt}
  \mid
\hspace{-1.5pt}
   \mathbf{X}_\tau)
\hspace{-2pt}
 \approx
\hspace{-2pt}
  -
\hspace{-0.5pt}
  \nabla_{\mathbf{X}_\tau}
\hspace{-1pt}
   \tfrac{1}{\sigma_Y^2}
\hspace{-2pt}
  \left\|\mathbf{M} (\mathbf{Y}
\hspace{-2pt}
   -
\hspace{-2pt}
    \hat{\mathbf{X}}_0(\mathbf{X}_\tau) )\right\|_F^2
\hspace{-2pt}
    ,
\label{eq:likelihood_gradient}
\end{equation}
where the gradient operator $\nabla_{\mathbf{X}_\tau}$ is implemented via automatic differentiation and therefore propagates through the score model $s_\theta(\mathbf{X}_\tau, \tau)$.
Because the score model is a neural network trained to exploit dependencies in the chosen representation, the backpropagated gradient is not confined to the range of $\mathbf{M}$ and also produces non-zero components in the null space of $\mathbf{M}$.
Consequently, the likelihood score guides not only the observed entries but also the missing region through correlations learned during training.
Combined with the prior score, it steers the sampling trajectory towards solutions consistent with the observations $\mathbf{Y}$.

\vspace{-3pt}
\section{Similarity-Guided Diffusion}
\vspace{-2pt}

\begin{figure}[t]
    \centering
     \resizebox{0.9\columnwidth}{!}{
    \hspace{-20pt}\input{figures/simdps_diagram}
    }
    \vspace{-10pt}
    \caption{Diagram of the proposed SimDPS framework.
    \vspace{-40pt}
    }
    \label{fig:diagram}
\end{figure}
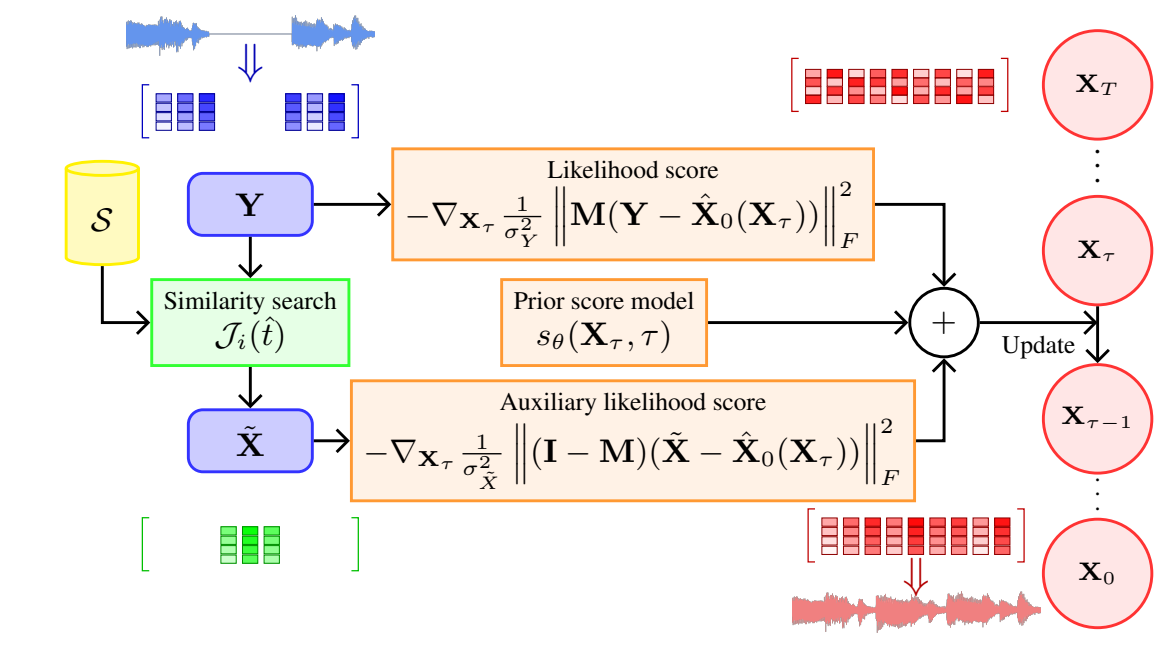

While posterior sampling is effective for short- to medium-length gaps \cite{moliner_diffusion-based_2024}, it becomes less reliable for longer gaps ($>300\,\mathrm{ms}$), where the generated content becomes more susceptible to contextual drift.
The likelihood score in \eqref{eq:likelihood_gradient} is defined only on the observed rows, so the missing portion---which lies in the null space of $\mathbf{M}$---is conditioned only indirectly through the score network and therefore remains weakly constrained.
As a result, the diffusion prior may produce reconstructions that are locally smooth and acoustically plausible yet still diverge from the expected musical content.

To address this limitation, we introduce an auxiliary representation $\tilde{\mathbf{X}} \in \mathbb{R}^{N \times C}$ that acts as a first approximation to the unknown clean sample $\mathbf{X}_0$ and contributes non-zero components in the null space of $\mathbf{M}$.
As described in Sec.~\ref{sec:sim}, this auxiliary representation is obtained through a similarity-search procedure that uses the observed context $\mathbf{Y}$ as reference.
The search can be performed either in waveform space or in latent space and may return either a single candidate or several high-scoring candidates.
The resulting auxiliary information is incorporated into the modified likelihood described in Sec.~\ref{sec:unilik}, thereby guiding the reverse diffusion process through a revised posterior.
An overview of the SimDPS framework is shown in Figure~\ref{fig:diagram}.

\vspace{-3pt}
\subsection{Modified Likelihood with Similarity Guidance}
\label{sec:unilik}
\vspace{-2pt}

We model the auxiliary guide as a noisy observation of the ground truth on the missing region:
\begin{equation} \label{eq:approx}
   (\mathbf{I} - \mathbf{M})\tilde{\mathbf{X}} = (\mathbf{I} - \mathbf{M})(\mathbf{X} + \boldsymbol{\varepsilon}_{\tilde{X}}),
\end{equation}
where $\boldsymbol{\varepsilon}_{\tilde{X}}$ captures the uncertainty of the auxiliary variable and is assumed to have i.i.d. Gaussian entries with variance $\sigma_{\tilde{X}}^2$.
We then construct a synthetic measurement
\begin{equation}
    \tilde{\mathbf{Y}} = \mathbf{M}\mathbf{Y} + (\mathbf{I} - \mathbf{M})\tilde{\mathbf{X}},
    \label{eq:synthetic_measurement}
\end{equation}
which combines the observed measurement $\mathbf{Y}$ with the auxiliary estimate $\tilde{\mathbf{X}}$ projected into the null space of $\mathbf{M}$.
Substituting \eqref{eq:inverse} and \eqref{eq:approx} yields $\tilde{\mathbf{Y}} = \mathbf{X} + \boldsymbol{\varepsilon}$ in the corresponding surrogate model, where $\boldsymbol{\varepsilon}$ is a non-isotropic Gaussian random matrix that captures both observed-data and auxiliary uncertainty.
More precisely, row $i$ of $\boldsymbol{\varepsilon}$ has covariance $\left[m_i \sigma_Y^2 + (1-m_i)\sigma_{\tilde{X}}^2\right]\mathbf{I}_C$.
Because the observed and auxiliary terms act on complementary subspaces, the synthetic log-likelihood decomposes into two additive terms:
$\log p(\tilde{\mathbf{Y}} \mid \mathbf{X}) = \log p(\mathbf{Y} \mid \mathbf{X}) + \log p(\tilde{\mathbf{X}} \mid \mathbf{X})$.

The modified likelihood integrates directly into \eqref{eq:bayes_score_decomp}.
Defining
$\mathcal{L}_Y(\mathbf{X}_\tau)\hspace{-3pt}=\hspace{-3pt}\left\| \mathbf{M}(\mathbf{Y} \hspace{-2pt}- \hspace{-2pt} \hat{\mathbf{X}}_0(\mathbf{X}_\tau)) \right\|_F^2\hspace{-2pt}$
and
$\mathcal{L}_{\tilde{X}}(\mathbf{X}_\tau)\hspace{-2pt}=\hspace{-2pt}\left\| (\mathbf{I}\hspace{-2pt}-\hspace{-2pt}\mathbf{M})(\tilde{\mathbf{X}} \hspace{-2pt}- \hspace{-2pt}\hat{\mathbf{X}}_0(\mathbf{X}_\tau)) \right\|_F^2$,
the modified likelihood score can be approximated analogously to \eqref{eq:likelihood_gradient} as
\begin{equation}
\nabla_{\mathbf{X}_\tau}
\hspace{-2pt}
\log p_\tau(\tilde{\mathbf{Y}}
\hspace{-2pt}
\mid
\hspace{-2pt}
\mathbf{X}_\tau)
\hspace{-2pt}
\approx
\hspace{-2pt}
 -\nabla_{\mathbf{X}_\tau}
\hspace{-2pt}
\left(\tfrac{1}
    {\sigma_Y^2}
    \mathcal{L}_Y(\mathbf{X}_\tau)
\hspace{-2pt}
+
\hspace{-2pt}
\tfrac{1}{\sigma_{\tilde{X}}^2}\mathcal{L}_{\tilde{X}}(\mathbf{X}_\tau)
\hspace{-2pt}
\right).
\label{eq:unified_gradient}
\end{equation}
This yields a modified guidance mechanism in which the contributions of the observation $\mathbf{Y}$ and the auxiliary guide $\tilde{\mathbf{X}}$ are weighted according to their respective uncertainties.
We parameterise the effective variances following \cite{moliner2023solving, shen2024understanding} as
$\sigma_Y^2 = \tfrac{\sigma(\tau)}{\omega_Y \sqrt{N C}} \left\| \nabla_{\mathbf{X}_\tau} \mathcal{L}_Y(\mathbf{X}_\tau) \right\|_F$,
with $\sigma_{\tilde{X}}^2$ defined analogously using a parameter $\omega_{\tilde{X}}$.
The weights $\omega_Y$ and $\omega_{\tilde{X}}$ control the balance between observed and auxiliary constraints and may be set to zero to disable either contribution.

\vspace{-3pt}
\subsubsection{Guidance from multiple candidates} \label{sec:multi}
\vspace{-2pt}

Beyond conditioning on a single inpaint candidate $\tilde{\mathbf{X}}$, we generalise the guidance by considering a set of the top-$K$ candidates, $\{(\tilde{\mathbf{X}}_i, \tilde{s}_i)\}^K_{i=1}$, ranked by their similarity scores $\tilde{s}_i$. We represent this set as a discrete weighted empirical distribution $p_{\mathrm{cand}}(x) = \sum^K_{i=1}  \sigma(\tilde{\mathbf{s}})_i \delta(x-\tilde{\mathbf{X}}_i),$ where $ \sigma(\tilde{\mathbf{s}})_i$ denotes the softmax-normalised confidence weight for the $i$-th candidate.

Extending \eqref{eq:unified_gradient}, we incorporate the influence of the entire ensemble via a weighted multi-candidate loss:
\begin{equation}
\textstyle
\mathcal{L}_{\tilde{X}}
\hspace{-1pt}
(\mathbf{X}_\tau)
\hspace{-2pt}
=
\hspace{-2pt}
\sum^K_{i=1}
\hspace{-2pt}
\sigma(\tilde{\mathbf{s}})_i \
 d_\mathrm{spec}(
\hspace{-1pt}
\tilde{\mathbf{X}}_i,
\hspace{-1pt}
\hat{\mathbf{X}}_0(\mathbf{X}_\tau)
\hspace{-1pt}
).
\end{equation}
Instead of employing a Frobenius norm across multiple candidates, which we observe produces conflicting gradients, we opt for a spectral loss $d_\mathrm{spec}$ tailored to this setup.

Our motivation is to emphasise the macroscopic structural similarities while discarding high-frequency variations.
We achieve this by mapping candidates to the frequency domain along the temporal dimension:
$\psi_\mathbf{X} = \mathcal{F}( (\mathbf{I} - \mathbf{M}) \odot\mathbf{X})$
and isolating the low-frequency content by applying a low-pass filter $\mathcal{H}_\mathrm{LP}$, with cutoff frequency $f_c$.
The spectral loss is defined as:
\begin{equation}
 d_\mathrm{spec}(\tilde{\mathbf{X}}, \hat{\mathbf{X}}_0(\mathbf{X}_\tau))
 \hspace{-2pt}
 =
 \hspace{-2pt}
 \big\lVert |\psi_{\tilde{\mathbf{X}}} \odot \mathcal{H}_\mathrm{LP}| - |\psi_{\hat{\mathbf{X}}_0(\mathbf{X}_\tau)} \odot \mathcal{H}_\mathrm{LP}| \big\rVert^2_F  .
\end{equation}

Inspired by \cite{falck2025fourier}, we restrict the multi-candidate guidance to early inference stages ($\sigma(\tau) \geq \sigma_\mathrm{multi}$). This acts as an ensemble search signal regulariser, guiding generation toward a low-frequency manifold defined by the top-$K$ candidates. At later steps, candidate guidance is removed, allowing the pre-trained prior to independently refine high-frequency details. We refer to the multiple candidate version of our method as \textit{SimDPS-Multi}, while the single-candidate version is referred to as  \textit{SimDPS-Solo}.

\vspace{-3pt}
\subsection{Similarity Search} \label{sec:sim}
\vspace{-2pt}

Inspired by \cite{perraudin_inpainting_2018}, we construct the auxiliary guide $\tilde{\mathbf{X}}$ by searching a source corpus $\mathcal{S} = \{\mathbf{S}_i \in \mathbb{R}^{L_i \times C}\}_{i=1}$ for a segment with surrounding context matching that of the observation $\mathbf{Y}$. Each source signal $\mathbf{S}_i$ may have arbitrary length $L_i > N$ and $C$ channels. While \cite{perraudin_inpainting_2018} retrieves variable-length segments, we extract fixed length-$N$ candidates matching the observation window, enabling temporally consistent evaluation across various methods.

To evaluate a  candidate $\mathbf{S}_{i,\hat{t}}$, starting at index $\hat{t}$ in source $\mathbf{S}_i \in \mathcal{S}$,
we compute the similarity cost over the context:
\begin{equation}
\mathcal{J}_i(\hat{t}) = \sum_k  d_k\big(\mathbf{Y},\, \mathbf{M} \mathbf{S}_{i,\hat{t}}\big),
\end{equation}
where $d_k(\cdot, \cdot)$ are domain-specific distance metrics. Candidate positions overlapping the masked interval are discarded.
The single best candidate is selected according to
\begin{equation}\label{eq:candidate_pair}
(\mathbf{S}^*, \hat{t}^*) = \arg \min_{\mathbf{S}_i \in \mathcal{S},\, \hat{t}} \mathcal{J}_i(\hat{t}), \quad \tilde{\mathbf{X}} = \mathbf{S}^*_{\hat{t}^*},
\end{equation}
yielding a similarity score $s = 1/(1 + \mathcal{J}_{*}(\hat{t}^*))$.

We now outline the domain-specific distance metrics and search constraints employed for latent representations and waveforms, respectively. Across both, contextual alignment close to the missing interval is emphasised by applying a temporal weight matrix $\mathbf{W}_k$,
with dimensions matching the $k$-th metric space. Using a metric-dependent context length,  $L_{c}$, the temporal weight ramps linearly from $0$ at the context extremes ($t_s - L_c$ and $t_e + L_c$) to $1$ at the gap boundaries ($t_s$ and $t_e$).

\vspace{-3pt}
\subsubsection{Latent-domain search}
\vspace{-2pt}

When the diffusion prior operates in the latent domain, both the observation $\mathbf{Y}$ and the source corpus $\mathcal{S}$ are represented in the same autoencoder space as used by the generative model.
Search is performed along the temporal axis; because these sequences have low temporal resolution (approximately 25 Hz in our setting), an exhaustive search remains computationally efficient.
A single distance metric ($k=1$), the weighted squared Frobenius distance, is used to compare the contexts of each candidate to the observation:
\begin{equation}
    d(\mathbf{Y},\mathbf{M} \mathbf{S}_{i,\hat{t}}) = \|\sqrt{\mathbf{W}_1} \odot (\mathbf{Y} - \mathbf{M} \mathbf{S}_{i,\hat{t}})\|_F^2.
\end{equation}

We propose two variants of the latent-space implementation: \emph{Solo} and \emph{Multi}.
In the \emph{Solo} variant, the guide $\tilde{\mathbf{X}}$ is formed from the single best-scoring candidate.
The \emph{Multi} variant retains several top-ranked candidates and calculates the combined guidance as described in Sec.~\ref{sec:multi}, aiming to reduce sensitivity to an imperfect single match.

\vspace{-5pt}
\subsubsection{Waveform-domain search}
\vspace{-3pt}

In the waveform setting, we
operate directly on audio samples.
Each distance metric, $d_k$, computes a weighted Frobenius distance over a feature domain:
$d_k(\mathbf{Y},\mathbf{M} \mathbf{S}_{i,\hat{t}}) = \big\| \sqrt{\mathbf{W}_k} \odot \big( \Phi_k(\mathbf{Y}) - \Phi_k(\mathbf{M} \mathbf{S}_{i,\hat{t}}) \big) \big\|^2_F$.
Here, $\{\Phi_k\}_{k=1}^K$ are feature mappings, and $\{d_k\}_{k=1}^K$ is a set of distances designed to compare such features.

Because exhaustive sample-accurate search is computationally expensive, we employ a two-stage procedure. First, we evaluate the search objective at a lower temporal resolution to identify an initial candidate. We then locally refine this estimate within a small window by minimising the sample-level continuity error at the gap boundaries.

\vspace{-5pt}
\section{Experiments and Results}
\vspace{-3pt}

We evaluate SimDPS across two distinct settings.
The first employs an \ac{ldm} pre-trained on general music, which we assess using both objective and subjective metrics.
We also report an additional setting based on a waveform-domain diffusion model specialised on piano recordings.

\vspace{-5pt}
\subsection{Latent Diffusion on General Music}
\vspace{-3pt}

We base our experiments on the open-source music foundation model ACE-Step 1.5 \cite{gong2026ace}, a \ac{ldm} trained on a large-scale music dataset. This setting allows us to evaluate \textit{SimDPS} in a broad, non-domain-specific scenario.

Only two components are used from ACE-Step 1.5: the autoencoder and the Diffusion Transformer backbone.
The autoencoder is a waveform-domain VAE \cite{evans2025stable} with a temporal rate of 25\,Hz and 64 channels.
For the Diffusion Transformer, we use the public ``base'' checkpoint from the generative pre-training stage, which is trained with the flow-matching objective to predict a velocity field, denoted $v_\theta$ following \cite{lipman2022flow}. This velocity estimate can be converted to a score estimate \cite{schusterbauer2025diff2flow}. Under the diffusion-time convention adopted in this paper, the conversion is given by:
\begin{equation}
   s_\theta(\mathbf{X}_\tau, \tau) = \tfrac{-1}{\tau^2 + \tau} \left(v_\theta\left(\tfrac{1}{\tau+1}\mathbf{X}_\tau, \tfrac{\tau}{\tau+1}\right)+ \mathbf{X}_\tau \right).
\end{equation}
Furthermore, because the model is trained with condition dropout, a common practice when using Classifier-Free Guidance \cite{ho2022classifier}, we can evaluate it without prompts or other conditioning inputs, effectively using it as an unconditional diffusion model.
We also observed that the model tends to end generated samples with silence or ending-like musical material, so we generate 5 extra seconds that get later discarded to prevent this bias from affecting the results.

Across all latent methods, we run the inference schedule over $T=100$ time steps.
We adopt the second-order stochastic sampler from \cite{karras2022elucidating}, setting the schedule parameters as $\sigma_\mathrm{max} = 20$,  $\sigma_\mathrm{min} = 10^{-3}$,  $\rho=7$,
and stochasticity parameter $S_\mathrm{churn}=10$.
We set the likelihood weighting parameter $\omega_Y=0.5$.
For \textit{SimDPS-Solo}, we set $\omega_{\tilde{X}} = 0.3$, and for \textit{SimDPS-Multi}, $\omega_{\tilde{X}} = 0.1$.
In \textit{SimDPS-Multi},
we only apply the similarity guidance until the minimum noise threshold $\sigma_\mathrm{multi}=0.5$ is passed, and set the spectral loss' cutoff frequency $f_c = 1.6$\,Hz.
We operate the latent search with a context length approximately equivalent to $1.0$\,s, and retain the top-$15$ candidates for \textit{SimDPS-Multi}.

\begin{table*}[t]
\centering
\caption{Objective evaluation metrics for the latent-domain setting. Best results are in \textbf{bold} and second-best are \underline{underlined}.
}
\label{tab:objective_metrics}
    \resizebox{0.95\textwidth}{!}{
\begin{tabular}{lcccccc}
\toprule
Method & MFCC dist. ($\downarrow$) & Chroma sim. ($\uparrow$) & Rhythm corr. ($\uparrow$) & Boundary err. ($\downarrow$) &MusicFM dist. ($\downarrow$)   &CLAP sim. ($\uparrow$)\\
\midrule
Latent-Interp.                  & $135.5 \pm 53.3$             & $0.924 \pm 0.054$            & $0.70 \pm 0.12$              & $ 2.57 \pm 1.69$             &$18.75 \pm 1.76$               &$0.610 \pm 0.116 $\\
Latent-Sim.                     & $112.1 \pm 64.7$             & $0.934 \pm 0.056$            & $0.70 \pm 0.17$              & $ 2.84 \pm 1.67$             &$17.13 \pm 2.32$               &$\underline{0.842} \pm 0.092 $\\
RePaint                        & $194.3 \pm 85.7$             & $0.887 \pm 0.078$            & $0.58 \pm 0.18$              & $ 2.82 \pm 1.43$             &$16.18 \pm 1.39$               &$0.626 \pm 0.140 $\\
DPS                            & $105.6 \pm 41.7$             & $0.928 \pm 0.055$            & $0.72 \pm 0.13$              & $ 1.98 \pm 0.75$             &$15.19 \pm 1.32$               &$0.823 \pm 0.087 $\\
SimDPS-Solo (proposed)       & $\underline{103.8} \pm 58.2$ & $\underline{0.936} \pm 0.053$& $\underline{0.73} \pm 0.13$     & $\underline{1.72} \pm 0.61$  &$\underline{14.46} \pm 1.33$               &$0.841 \pm 0.090 $\\
SimDPS-Multi (proposed)      & $\mathbf{97.8} \pm 50.9$     & $\mathbf{0.942} \pm 0.050$   & $\mathbf{0.75} \pm 0.13$  & $\mathbf{1.70} \pm 0.56$     &$\mathbf{14.42 \pm 1.34}$   &$\mathbf{0.843\pm 0.088}$\\
\bottomrule
\end{tabular}
}
\end{table*}

\begin{figure*}[!t]
    \centering
    \includegraphics[trim=7 11 6 7 , clip, width=0.88\textwidth]{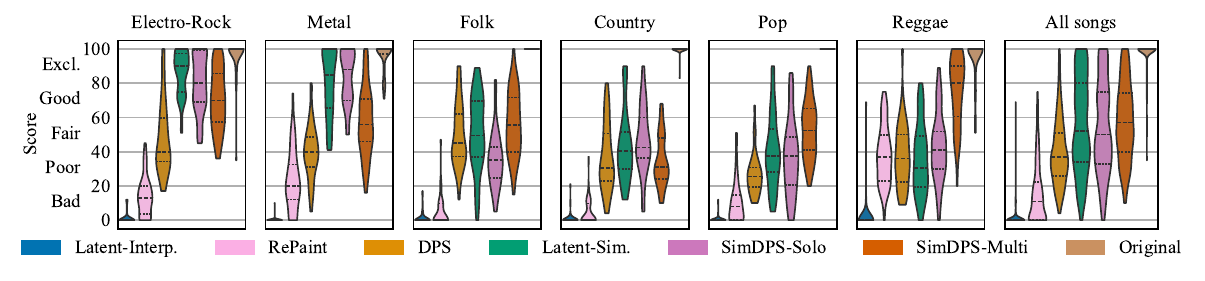}
    \vspace{-7pt}
    \caption{Subjective evaluation of latent-space methods on general music. Violin plots show score distributions per example and aggregated across the full dataset.
    \vspace{-5pt}
    }

    \label{fig:violins_latent}
\end{figure*}

\vspace{-5pt}
\subsubsection{Objective evaluation}
\vspace{-3pt}

Although exact recovery is not expected in long-gap inpainting, we argue that successful reconstructions should preserve the structural and stylistic properties of the original musical context.
We therefore design an objective evaluation, comparing the method reconstructions with the ground-truth references using a set of complementary feature-based metrics that assess different aspects of the reconstruction.

For this evaluation, we use 500 excerpts of length 15\,s from the test split of MUSDB18-HQ \cite{musdb18-hq}.
In each excerpt, a 5-s gap is placed at random within the segment's interior.
For all retrieval-based methods, the source corpus used for similarity search consists of the remainder of the same song outside the masked interval.

We compare six reconstruction conditions.
\textit{Latent-Interp} performs linear interpolation in the latent space and serves as a low anchor.
\textit{Latent-Sim} directly inserts the highest-scoring retrieved latent segment into the gap.
\textit{DPS} denotes unguided diffusion-based inpainting, following \cite{moliner_diffusion-based_2024}.
We further evaluate the two variants of the proposed method introduced in Sec.~\ref{sec:sim}: \textit{SimDPS-Solo}, which uses the single best-matching candidate, and \textit{SimDPS-Multi}, which combines guidance from multiple high-scoring candidates.
Finally, for completeness, we include the supervised inpainting module provided in the ACE-Step codebase, denoted \textit{RePaint} as inspired by an image inpainting method ~\cite{lugmayr2022repaint}.

We report six complementary reference-based metrics.
\emph{MFCC distance} measures timbral consistency using perceptually-motivated spectral features. \emph{Rhythm correlation} is defined as the cross-correlation between onset strength envelopes and captures coarse rhythmic agreement.
\emph{Chroma cosine similarity}  \cite{copet2023simple} evaluates harmonic agreement while being relatively insensitive to fine timbral variation.
These three metrics compare only the audio content within the reconstructed gap.

To assess how well the reconstructed segment connects to the surrounding audio, we additionally compute a \emph{boundary error}, defined as the log-spectral distance to the reference in short analysis windows centred at the gap boundaries. Finally, we include two embedding-based distances computed on the full excerpt rather than on the gap. \emph{CLAP cosine similarity} provides a coarse measure of semantic similarity, whereas \emph{MusicFM Euclidean distance} measures discrepancy in a representation space shown to capture higher-level musical structure \cite{won2024foundation,toyama2025foundational}. Using the full excerpt is important for these embedding metrics because the representations are more informative when both the reconstructed region and its surrounding context are available.

The results of the objective evaluation are displayed in Table \ref{tab:objective_metrics}.
The \textit{RePaint} method performs poorly across most metrics, which we attribute to an out-of-distribution context length (15\,s vs. full-length songs). This failure highlights how sensitive supervised models are when deployed outside their training regime.
\textit{SimDPS-Multi} achieves the greatest score across all metrics reported.
\textit{SimDPS-Multi} also achieves a notably better MFCC distance than each of the other diffusion methods, suggesting a benefit to multi-candidate guidance.
There is notable variance in the results of both \textit{Latent-Sim} and \textit{DPS} across metrics.
In particular, \textit{Latent-Sim} performs poorly on the Boundary Error and MusicFM embedding, with a markedly higher standard deviation than all other methods. This is an example of the weakness of deterministic similarity methods, in cases where the best match available is still poor.
\textit{DPS} performs comparatively well here with the lowest standard deviation, yet receives a poorer result on Chroma cosine similarity, CLAP and Rhythm correlation than \textit{Latent-Sim}; indicating---on average---the context adherence issues that can arise with unguided long-gap DPS inpainting. The proposed \textit{SimDPS} methods perform consistently across each baseline, demonstrating the benefit to combine traditional similarity corpus search with probabilistic generative methods.

\vspace{-5pt}
\subsubsection{Subjective evaluation} \label{sec:subjective_latent}
\vspace{-3pt}

The same set of conditions is also evaluated in a subjective listening test.
We adopt a multi-stimulus listening test in which listeners assess the perceptual plausibility of the reconstructions.
The test was implemented with webMUSHRA \cite{schoeffler2018webmushra}.
Six musical excerpts of varying genres are selected from songs in the MUSDB18-HQ \cite{musdb18-hq} test split.
In each trial, the masked excerpt is provided as a reference for the surrounding musical context. Listeners are asked to evaluate the six reconstruction conditions according to how plausible they sound amongst the context.
A total of 23 volunteers participated in the study.

\begin{figure*}[t!]
    \centering
    \includegraphics[trim=7 11 6 7 , clip, width=0.88\textwidth]{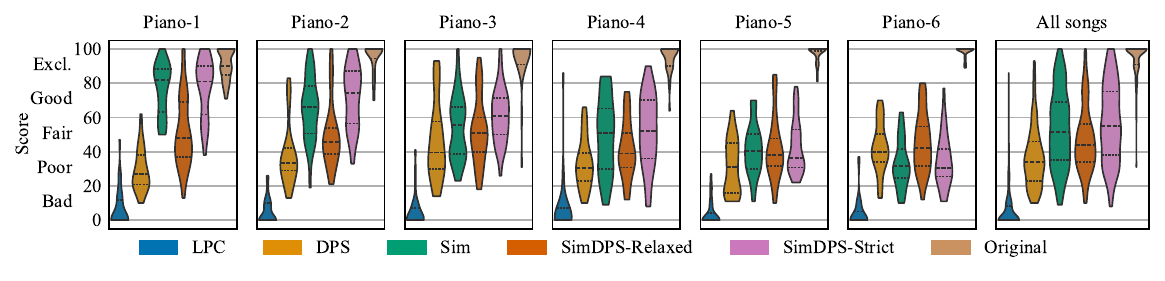}
    \vspace{-7pt}
    \caption{Subjective evaluation of waveform-domain methods on piano music. Violin plots show score distributions per example and aggregated across the full dataset.
    \vspace{-7pt}
    }

    \label{fig:boxplots}
\end{figure*}

Figure \ref{fig:violins_latent} shows the results of the listening experiment.
All participants generally identified both the reference and anchor.
The experiment confirms the same broad trend observed in the objective evaluation: \textit{SimDPS} is generally preferred over the \textit{DPS} and \textit{RePaint} baselines, while more nuanced patterns emerge when comparing the two versions of \textit{SimDPS} with \textit{Latent-Sim}.

The responses are strongly example-dependent and, in particular, depend on the quality of the retrieved similarity match.
For excerpts where the best match is already highly coherent, such as \textit{Electro-Rock} and \textit{Metal}, the \textit{Latent-Sim} baseline performs very well and leaves limited room for further improvement. In these cases, \textit{SimDPS-Solo} remains close to \textit{Latent-Sim} and does not noticeably degrade performance, whereas \textit{SimDPS-Multi} can be detrimental. By contrast, when the retrieved match is weaker, \textit{SimDPS-Multi} becomes more advantageous: in \textit{Reggae} and \textit{Pop}, \textit{SimDPS-Multi} can demonstrate robustness to a poorly rated similarity match by providing significant improvements over \textit{Latent-Sim} ($p < 0.01$ and $p = 0.02$, respectively\footnote{All $p$-values reported are derived from pairwise comparisons using a Wilcoxon signed-rank test.}), while in \textit{Folk} the improvement is more marginal ($p=0.12$).
\textit{Country}, however, is an exception to this trend, as \textit{SimDPS-Multi} achieves an equivalent score to the unguided \textit{DPS} baseline, improved upon by \textit{SimDPS-Solo}.

\vspace{-5pt}
\subsection{Waveform Diffusion on Piano}
\vspace{-3pt}

As an additional evaluation, we employ a waveform-domain diffusion model trained on piano recordings to determine whether our guidance scheme is also compatible with waveform-domain priors.

For the backbone score model $s_\theta$, we use a waveform-domain diffusion model with the MR-CQTdiff architecture  \cite{dacosta2025mrcqtdiff}.
The model is trained on the MAESTRO dataset \cite{hawthorne_enabling_2019} using mono recordings at 44.1\,kHz and training segments of 6\,s.
Initial training was performed on the full train split for 180k iterations, followed by 320k iterations on recordings from 2017–2018, which have superior acoustic quality.

To sample from the diffusion model, we discretise the time $\{\tau_i\}_{i=1}^T$ into $T=50$ steps.
We set a maximum $\tau_T=8$ and a minimum $\tau_1=e^{-5}$, using
a logarithmic discretisation schedule.
 We employ the second-order stochastic sampler proposed by Karras et al.~\cite{karras2022elucidating},
 setting the stochasticity parameter $S_{\mathrm{churn}}=10$.

We observe that the hyperparameter $\omega_{\tilde{X}}$, which modulates the strength of the similarity-based guidance, is critical to waveform inpainting performance. To evaluate this impact, we define two configurations of our method: \textit{SimDPS-Strict} ($\omega_{\tilde{X}} = 0.15$) and \textit{SimDPS-Relaxed} ($\omega_{\tilde{X}} = 0.04$). For both versions, the reconstruction weighting is held constant at $\omega_Y = 0.3$.
Notably, the multiple candidate configuration \textit{SimDPS-Multi} is omitted from this experiment because initial testing indicated it was ineffective at generating adequate samples in this context.

To perform the feature-based similarity search, the audio signals are resampled to $12$\,kHz, improving search efficiency.
The source corpus $\mathcal{S}$ comprises the complete segments of the music track preceding and following the gap.
We selected two search features: the STFT (0.75\,s context) to maintain short-range consistency and a chromagram (5.0\,s context) to capture long-range musical coherence \cite{muller_music_2021}.
For both, we used a Hann window of length 1024 and hop size 256.
This allowed efficient joint frame pre-processing.

\vspace{-5pt}
\subsubsection{Subjective evaluation}
\vspace{-3pt}

We conduct a listening test following the same protocol as described in Sec.~\ref{sec:subjective_latent}. In this case, each trial featured a 6-s piano excerpt with a 2-s central gap, where participants compared the original reference against five reconstruction methods. Although 17 volunteers participated, one was excluded from the final analysis due to unreliable responses.

Among the evaluated methods, Linear Predictive Coding (\textit{LPC}) \cite{kauppinen_audio_2002} served as a low anchor, as it is generally unsuitable for gaps of this duration.
The \textit{Sim} method directly copies the similarity match, adhering to the original method \cite{perraudin_inpainting_2018}.
We also included \textit{DPS}, which refers to the method by \cite{moliner_diffusion-based_2024}, alongside two variants of our proposed approach, \textit{SimDPS-Relaxed} and \textit{SimDPS-Strict}.

The listening test results,  shown in Figure~\ref{fig:boxplots}, indicate similar patterns as in the latent case.
 Across all excerpts, the reference (\textit{Original}) was consistently rated Excellent, while the anchor (\textit{LPC}) remained in the Poor range.

 We analyse ratings on a per-excerpt basis.
The similarity baseline (\textit{Sim}) is a useful proxy for match quality. When the retrieved segment was highly coherent, as in \textit{Piano-1}, \textit{SimDPS-Strict} matched \textit{Sim}; when the match was only moderate, it improved on \textit{Sim}, significantly for \textit{Piano-2} and \textit{Piano-3} (both $p = 0.02$) and marginally, while not statistically significant, for \textit{Piano-4} ($p = 0.2$).
This shows that employing a diffusion model can improve upon moderately fitting matches.
When the retrieved match was poor, however, \textit{SimDPS-Strict} offered no consistent advantage.

The unguided \textit{DPS} condition was generally rated low, underscoring the importance of guidance, while \textit{SimDPS-Relaxed} usually improved on \textit{DPS} but remained below or equal to \textit{Sim}. \textit{Piano-6} was the main exception, where \textit{DPS} and \textit{SimDPS-Relaxed} outscored both \textit{Sim} and \textit{SimDPS-Strict}, although all methods still received low ratings. Overall, \textit{SimDPS-Strict} achieved the highest mean score.

\vspace{-7pt}
\section{Conclusions}
\vspace{-3pt}

This paper introduces SimDPS, a training-free and model-agnostic framework for long-gap music inpainting that combines similarity retrieval with diffusion posterior sampling.
Using an auxiliary guide, it applies a non-uniform posterior covariance matrix to direct samples toward plausible, context-aligned solutions.
SimDPS operates in both latent and waveform domains, and in latent space can additionally exploit multiple high-scoring candidates through \textit{SimDPS-Multi}.
Using an open-source latent diffusion model, SimDPS achieved the highest objective scores, improving upon unguided inpainting and pure similarity search without any model fine-tuning or decoder passes.
Subjective evaluations show that SimDPS significantly enhances perceptual plausibility, frequently outperforming standalone similarity
search in cases where retrieval candidates are suboptimal.
Variable success across samples is the main limitation, and studies into creative inference applications over varied corpora warrant further investigation.

\newpage

\section{Acknowledgment}
This work was supported by Nokia Solutions and Networks, the Research Council of Finland (Grant no.~371845, REMUS---Restoring Music Recordings Using Generative Models), and the HUCE infrastructure of the Aalto School of Electrical Engineering. We acknowledge the computational resources provided by the Aalto University “Science-IT” project.
We thank the volunteers who participated in the listening tests.

\bibliography{ISMIRtemplate}

\end{document}

%% file: figures/simdps_diagram.tex
\newcommand{\latenticonunit}{0.19cm}
\newcommand{\waveformpreviewwidth}{2.40cm}
\newcommand{\scoreboxwidth}{4.75cm}

\newcommand{\latentstack}[3]{%
    \foreach \y/\delta in {0.68/-16,1.16/-5,1.64/5,2.12/16} {
        \pgfmathtruncatemacro{\rowshade}{max(4,min(96,#3+\delta))}
        \fill[#1!\rowshade] (#2,\y) rectangle ++(0.78,0.42);
        \draw[#1!55!black, line width=0.15pt] (#2,\y) rectangle ++(0.78,0.42);
    }
}

\newcommand{\maskedlatenticon}[1]{%
\tikz[baseline=(current bounding box.center), x=\latenticonunit, y=\latenticonunit]{%
    \draw[#1!75!black, line width=0.35pt] (0.35,0.30) -- (0,0.30) -- (0,3.00) -- (0.35,3.00);
    \draw[#1!75!black, line width=0.35pt] (10.65,0.30) -- (11,0.30) -- (11,3.00) -- (10.65,3.00);
    \foreach \x/\shade in {0.7/18,1.8/42,2.9/68,7.3/60,8.4/24,9.5/76} {
        \latentstack{#1}{\x}{\shade}
    }
}}

\newcommand{\gaplatenticon}[1]{%
\tikz[baseline=(current bounding box.center), x=\latenticonunit, y=\latenticonunit]{%
    \draw[#1!75!black, line width=0.35pt] (0.35,0.30) -- (0,0.30) -- (0,3.00) -- (0.35,3.00);
    \draw[#1!75!black, line width=0.35pt] (10.65,0.30) -- (11,0.30) -- (11,3.00) -- (10.65,3.00);
    \foreach \x/\shade in {4.0/36,5.1/82,6.2/48} {
        \latentstack{#1}{\x}{\shade}
    }
}}

\newcommand{\fulllatenticon}[1]{%
\tikz[baseline=(current bounding box.center), x=\latenticonunit, y=\latenticonunit]{%
    \draw[#1!75!black, line width=0.35pt] (0.35,0.30) -- (0,0.30) -- (0,3.00) -- (0.35,3.00);
    \draw[#1!75!black, line width=0.35pt] (10.65,0.30) -- (11,0.30) -- (11,3.00) -- (10.65,3.00);
    \foreach \x/\shade in {0.7/18,1.8/42,2.9/68,4.0/36,5.1/82,6.2/48,7.3/60,8.4/24,9.5/76} {
        \latentstack{#1}{\x}{\shade}
    }
}}

\newcommand{\noiselatenticon}[1]{%
\tikz[baseline=(current bounding box.center), x=\latenticonunit, y=\latenticonunit]{%
    \draw[#1!75!black, line width=0.35pt] (0.35,0.30) -- (0,0.30) -- (0,3.00) -- (0.35,3.00);
    \draw[#1!75!black, line width=0.35pt] (10.65,0.30) -- (11,0.30) -- (11,3.00) -- (10.65,3.00);
    \foreach \x/\sA/\sB/\sC/\sD in {
        0.7/86/18/67/35,
        1.8/28/79/22/91,
        2.9/73/44/88/16,
        4.0/52/24/77/63,
        5.1/14/95/38/82,
        6.2/69/31/57/20,
        7.3/41/84/26/72,
        8.4/92/36/61/13,
        9.5/23/75/47/89
    } {
        \foreach \y/\shade in {0.70/\sA,1.14/\sB,1.58/\sC,2.02/\sD} {
            \fill[#1!\shade] (\x,\y) rectangle ++(0.78,0.42);
            \draw[#1!55!black, line width=0.15pt] (\x,\y) rectangle ++(0.78,0.42);
        }
    }
}}

\begin{tikzpicture}[
    font=\small,
    node distance=0.8cm and 0.5cm,
    blue_box/.style={
        rectangle, rounded corners, draw=blue!80, fill=blue!20, thick,
        minimum width=1.2cm, minimum height=0.6cm, text centered
    },
    orange_box/.style={
        rectangle, draw=orange!80, fill=orange!10, thick,
        minimum width=1.8cm, minimum height=0.6cm, text centered
    },
    green_box/.style={
        rectangle, draw=green!80, fill=green!10, thick,
        minimum width=1.8cm, minimum height=0.6cm, text centered
    },
    red_circle/.style={
        circle, draw=red!80, fill=red!10, thick,
        minimum size=1.05cm, text centered
    },
    sum_circle/.style={
        circle, draw=black, fill=white, thick, minimum size=0.2cm
    },
    data_cyl/.style={
        cylinder, shape border rotate=90, aspect=0.3, draw=yellow!100, fill=yellow!30, thick,
        minimum height=1cm, minimum width=0.7cm
    },
    latent_vis/.style={
        minimum width=2cm, minimum height=0.62cm,
        inner sep=0pt, outer sep=0pt
    },
    waveform_preview/.style={inner sep=0pt, outer sep=0pt, align=center},
    score_box/.style={orange_box, minimum width=\scoreboxwidth, inner xsep=5pt, inner ysep=3pt},
    codec_icon/.style={inner sep=0pt, outer sep=0pt},
    arrow/.style={thick, -Straight Barb},
    dashed_arrow/.style={thick, -Straight Barb, dashed}
]

\coordinate (cj) at (0,0);
\coordinate (cxT) at (8.2,2.3);
\coordinate (cupdate) at (8.2,0);

    \node (j) at (cj) [green_box, align=center] {\scriptsize Similarity search \\ $\mathcal{J}_i(\hat{t})$};

    \node (data) [data_cyl, above left=0.2cm and 0.1cm of j] {$\mathcal{S}$};
    \node (y)    [blue_box, above=0.4cm of j] {$\mathbf{Y}$};
    \node (ywaveform)  [above=0.2cm of y, align=center] {
    \scriptsize \maskedlatenticon{blue}};
    \node (yorigwave) [waveform_preview, above=0.15cm of ywaveform] {
    \includegraphics[width=\waveformpreviewwidth]{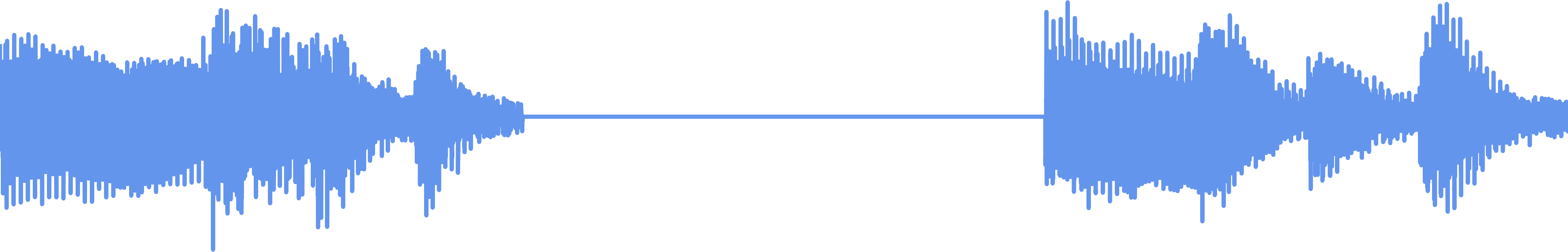}};

    \node (xtilde) [blue_box, below=0.4cm of j] {$\tilde{\mathbf{X}}$};
    \node (xtildewaveform)  [below=0.3cm of xtilde, align=center] {\scriptsize \gaplatenticon{green}};

    \node (prior)      [orange_box, right=1.45cm of j, align=center] {\scriptsize Prior score model \\ $s_\theta(\mathbf{X}_\tau, \tau)$};
    \node (likelihood) [orange_box, right=0.75cm of y, align=center] {\scriptsize Likelihood score \\
    $-\nabla_{\mathbf{X}_\tau} \frac{1}{\sigma_Y^2}\left\| \mathbf{M}(\mathbf{Y} - \hat{\mathbf{X}}_0(\mathbf{X}_\tau)) \right\|_F^2$};
    \node (aux)        [orange_box, right=0.35cm of xtilde, align=center] {\scriptsize Auxiliary likelihood score \\
    $-\nabla_{\mathbf{X}_\tau} \frac{1}{\sigma^2_{\tilde{X}}}\left\| (\mathbf{I}-\mathbf{M})(\tilde{\mathbf{X}} - \hat{\mathbf{X}}_0(\mathbf{X}_\tau)) \right\|_F^2$};

    \node (sum1)      [sum_circle, right=5.4cm of j] {$+$};

    \node (xT)  at (cxT)    [red_circle] {\scriptsize $\mathbf{X}_T$};
    \node (noise)  [left=0.2cm of xT] {\scriptsize \noiselatenticon{red}};
    \node (dots_l) [below=-0.25cm of xT] {$\vdots$};
    \node (xtau)   [red_circle, below=0cm of dots_l] {\scriptsize $\mathbf{X}_{\tau}$};

    \node (xtau-1) [red_circle, below=0.55cm of xtau] {\scriptsize $\mathbf{X}_{\tau-1}$};
    \node (dots_r) [below=-0.3cm of xtau-1] {\scriptsize $\vdots$};
    \node (x0)     [red_circle, below=-0.05cm of dots_r] {\scriptsize $\mathbf{X}_0$};
    \node (sample) [above left=-0.4cm and 0.2cm of x0] {\scriptsize \fulllatenticon{red}};
    \node (decodedwave) [waveform_preview, below=0.16cm of sample] {
    \includegraphics[width=\waveformpreviewwidth]{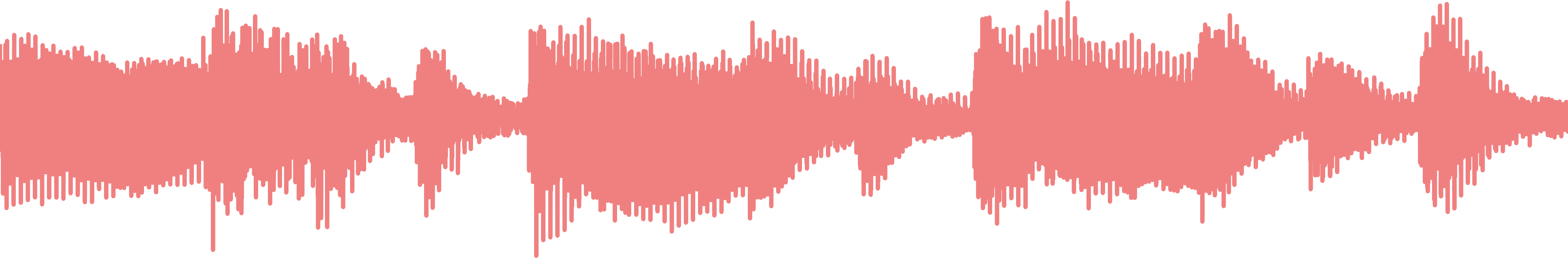}};

    \draw [arrow] (data) |- (j.west);
    \node [codec_icon, text=blue!70!black] at ($(yorigwave.south)!0.5!(ywaveform.north)$) {\large$\Downarrow$};
    \draw [arrow] (y.south) -- (j.north);
    \draw [arrow] (j.south) -- (xtilde.north);
    \draw [arrow] (xtilde.east) -- ++(0.3,0) |- (aux.west);
    \draw [arrow] (y) -- (likelihood.west);

    \draw [arrow] (likelihood.east) -| (sum1) node[midway, above, xshift=0.1cm, font=\scriptsize] {};
    \draw [arrow] (prior.east) -- (sum1);
    \draw [arrow] (sum1.east) -- (cupdate) node[midway, below, xshift=0cm, font=\scriptsize] {Update};
    \draw [arrow] (aux.east) -| (sum1) node[midway, below, xshift=0.1cm, font=\tiny] {};

    \draw [arrow] (xtau) -- (xtau-1);
    \node [codec_icon, text=red!70!black] at ($(sample.south)!0.1!(decodedwave.north)$) {\large$\Downarrow$};

\end{tikzpicture}